\documentclass[12pt]{article}
\usepackage{boldgreek}
\usepackage{amsmath}
\usepackage{amssymb}
\usepackage{epsf}
\newcommand{\eq}{$$} 
\newcommand{\beq}{\begin{equation}} 
\newcommand{\eeq}{\end{equation}\noindent}
\newcommand{\bear}{\begin{eqnarray*}}
\newcommand{\eear}{\end{eqnarray*}\noindent}
\newcommand{\bearn}{\begin{eqnarray}}
\newcommand{\eearn}{\end{eqnarray}\noindent}

\begin{document}
\begin{flushright}
LPT Orsay 00-72 \\
August 2000\\
\end{flushright}

\begin{center}
\vspace{24pt}

{\bf ON THE MICROSCOPIC DYNAMICS\\ OF DCC FORMATION} \\
\vspace{24pt}

Julien SERREAU
 
\vspace{10pt}

LPT, B\^{a}timent 210, Universit\'e Paris-Sud, 91405 Orsay, 
France\footnote{Laboratoire associ\'e au Centre National
de la Recherche Scientifique - URA00063.}
\vspace{10pt}

\begin{abstract}
The dynamics of the pion field after a quench is studied in the
framework of the linear sigma model. Our aim is to determine
to what extent the amplified pion field resembles the DCC
picture originally proposed in the early '90s. We present
the result of a computer experiment where, among other things, we
study in detail the correlation between isospin orientations
of the distinct modes of the field. We show that this correlation
is absent. In a sense, the distinct modes behave as distinct
DCCs. The implications of this observation are discussed. 
\end{abstract}
\end{center}
\vspace{15pt}

\section{Introduction}

A disoriented chiral condensate (DCC) is a medium where ``the quark 
condensate, $\langle 0 |q_L \bar q_R | 0 \rangle$, is chirally rotated 
from its usual orientation in isospin space"~\cite{bjorken}. This 
hypothetical object has attracted the attention of many physicists 
in the last decade and is still a topic of intense theoretical as well 
as experimental investigations (see reviews~\cite{reviews} and 
also~\cite{exp}). 
\par
One imagines a rapidly expanding fireball,
produced in a high energy hadronic or nuclear collision, whose interior 
is separated from the outside vacuum by a thin, hot shell of
hadronic debris. The quark condensate inside the bubble might then be 
misaligned with respect to its va\-cuum orientation. The DCC subsequently
decays toward ordinary vacuum by radiation of soft coherent pions.
The original picture of the DCC is that of a classical pion field
oscillating coherently in a well defined direction in isospin space~\cite{bj}
and indeed, Blaizot and Krzywicki showed that such field configurations can
develop with suitable initial conditions~\cite{bk}.
This phenomenological picture leads in particular to a very simple, but
very striking prediction: the $1/\sqrt f$ event-by-event distribution 
of the neutral fraction $f$ of radiated pions.
\par
The idea received a boost in 1993 when Rajagopal and Wilczek proposed
a microscopic mechanism for the formation of a classical coherent pion field 
inside the bubble~\cite{rw}, the so-called quench scenario: the rapid expansion
causes the system, initially thermalised above the critical temperature, to cool 
down very rapidly, leading to the amplification of the soft pion modes via the
mechanism of spinodal instability. Since then, this scenario has attracted much
attention and has been further developed. In particular, quantum effects in 
the mean field approximation have been included (see e.g.~\cite{boy,cooper}) 
and the drastic quench approximation has been abandoned in favour of one 
taking the expansion explicitly into account~\cite{cooper,ran}.
\par
The quench scenario has been widely accepted as a microscopic description
of DCC formation in heavy ion collisions and some works were dedicated to 
the study of its phenomenological implications~\cite{ggp,raj,ran2}. However
it is less clear than usually believed whether the typical field 
configuration emerging from a quenched
thermal ensemble is identical to what one had in mind in the early '90s.
Indeed the DCC configuration can be characterised by three essential 
features~\cite{bjorken,bj,bk}:
\begin{itemize}
\item[(a)]it is a coherent field excitation, essentially a classical field
configuration,
\item[(b)]each Fourrier component of the field oscillates along a 
  well defined orien\-tation in isospin space\footnote{In the following 
  we refer to the trajectory of one Fourier mode in isospin space as
  (iso-)polarisation. The case described in (b) will then be refered to as 
  linear polarisation.},
\item[(c)]the orientations of different modes are aligned along one 
single direction.
\end{itemize}
The object of the present paper is to perform a detailed statistical analysis
of the generic field configuration emerging from the quench scenario in order 
to determine whether it exhibits a DCC structure.
\par
For this purpose one has to choose a reliable framework.
Quantum effects are not of first importance as the system 
is essentially classical~\cite{bjorken}. Moreover these corrections
can, so far, only be included in a mean field approximation, not suitable
to study the correlations between modes. Finally, in a realistic scenario,
where the quench is due to a rapid expansion, the system enters only rarely 
in the instability region~\cite{krzjs}. Then, to keep the argument simple, 
it is worth going back to the original framework of Rajagopal and Wilczek 
in which the system, initially in a high temperature ($T>T_c$) configuration 
is evolved with the full, non-linear, zero temperature equations of the 
classical linear~$\sigma$-model.

\section{The formalism}

\subsection{The quench scenario}

The chiral field is parametrized by a four-component vector field
in chiral space: $\bfphi = (\bfpi,\sigma)$ and its time evolution
is governed by the classical linear~$\sigma$-model equations of motion
\beq
 \left( \partial^2  - \lambda v^2 + \lambda\bfphi^2 (\vec x,t) \right) 
 \bfphi (\vec x,t) = H \mbox{\bf n}_\sigma \, \, ,
 \label{eom}
\end{equation}\noindent
where the parameters $v$, $\lambda$ and $H$ are related to physical quantities 
via:
\eq
 m_\pi^2 = m_\sigma^2 - 2 \lambda f_\pi^2 =
 \lambda \left( f_\pi^2 - v^2 \right) \, , \, \,
 \mbox{and} \, \, H = f_\pi m_\pi^2 \, ,
\eq
and are chosen so that $m_\pi$=135 MeV, $m_\sigma$=600 MeV and $f_\pi$=92.5 MeV.
\par
The initial field configuration is sampled from
a thermal ensemble at some temperature $T > T_c$: the values
of $\bfphi$ and $\dot \bfphi$ are chosen independently on each site 
of the cubic lattice from gaussian distributions centered
around $\bfphi = \dot \bfphi = \bf 0$ and with 
variances\footnote{In ref.~\cite{rw}
  the variances are given for the lengths of the vectors $\bfphi$ and 
  $\dot \bfphi$,\\ 
  e.g. $\langle \bfphi^2 \rangle = 
  \sum_{j=1}^4\langle \phi_j^2 \rangle = v^2/4$.}
$\langle \phi_j^2 \rangle = v^2/16$ and $\langle \dot \phi_j^2 \rangle = v^2/4$
($j=1,2,3,4$).
The lattice spacing $a$ has then the physical interpretation of the 
degenerate  $\bfpi$ and $\sigma$ correlation length at temperature $T$, 
$a=(200$ MeV$)^{-1}$. 
\par
We denote by~$\bfvarphi (\vec k,t)$ and~$\dot \bfvarphi (\vec k,t)$ the 
Fourier components of the field and its time derivative at time~$t$, 
and choose Neumann boundary conditions so that these components are real. 
This is not an essential point, but it is more convenient to make this
choice in discussing the question of polarization. The infrared
cutoff in Fourier space is~$\Delta k = \pi / Na$ where~$Na$ is the length
of the cubic box. 
\par
The important point is elsewhere: the values of~$\bfphi$
at distinct lattice sites are assumed to be independant 
gaussian random numbers. One can easily convince oneself that this implies that
the values of the Fourier components~$\bfvarphi$ and~$\dot \bfvarphi$
 at distinct sites of the 
{\em discretized Fourier space} are {\em independent} gaussian random 
numbers as well. In other words, there is no correlation
between modes in the initial state. To create a DCC configuration, the quench
mechanism has not only to be efficient in amplifying the modes (which it is), 
but it should also be able to {\em build correlations} between amplified
modes.

\subsection{Observables}

Our goal is to determine whether a {\em generic} event in the above-described
statistical ensemble looks like an ideal DCC configuration. For this purpose 
we shall compute the event-by-event distribution of the neutral pion fraction 
$f(\vec k)$ in the mode~$\vec k$~(see~Eq.~(\ref{ratio})) on the one hand, and 
on the other hand we shall measure the correlations of this quantity between
different modes~(see~Eq.~(\ref{correl})). Let us briefly review the main ideas
underlying the physics we want to describe and argue for the relevance of the
above mentioned observables.
\par
In this simplified model, although the size of the box is fixed, we have in 
mind a rapidly expanding system (the quench assumption is a drastic 
idealization of the effect of expansion). This means that after some time
it becomes so dilute that the modes decouple and evolve freely.
To model this we stop the evolution at some freeze-out time 
$t_f$\footnote{A more
  realistic description would take into account the possible finite time extent 
  of the freeze-out period. This would require a model for freeze-out which is 
  out of the scope of this paper. We assume that each mode $\vec k$ freezes out 
  at $t_f$ which can be thought as the end of a common freeze-out period.}.
The field configuration at $t_f$ is the ``initial condition" 
for the subsequent free evolution and determines completely the properties 
of the outgoing waves which describe free propagating pions. Indeed,
one might consider the classical field configuration at $t_f$ as the
expectation value of the corresponding quantum field in the coherent 
state(we only consider the pion sector)
\beq
 | \alpha , t_f\rangle = \exp \left(\int d^3k \,\,
 \bfalpha(\vec k,t_f) \cdot \mbox{\bf a}^\dagger(\vec k) \right)|0\rangle \, ,
 \label{coherent}
\end{equation}\noindent
with
\eq
 \bfalpha \cdot \mbox{\bf a}^\dagger =
 \sum_{j=1}^3 \alpha_j a_j^\dagger \, ,
\eq
where~$a_j^\dagger(\vec k)$ is the creation operator of a free pion
with isospin~$j$ and momentum~$\vec k$, while~$\alpha_j(\vec k,t_f)$ is 
the eigenvalue of the corresponding annihilation operator ($t_f$ is a 
parameter). The 
state~(\ref{coherent}) is related to the field configuration at 
freeze-out through~($\omega_k = \sqrt{k^2 + m_\pi^2}$)
\beq
 \bfalpha(\vec k,t_f) = 
 \frac{i\dot \bfvarphi (\vec k,t_f) + \omega_k \bfvarphi (\vec k,t_f)}
 {\sqrt{2 \omega_k }} \, \, .
\label{alpha}
\end{equation}\noindent
The subsequent evolution reads\footnote{The "subsequent evolution" is 
  introduced for the purpose of the argument but is not essential. }
\beq
 \bfalpha(\vec k,t>t_f) = \bfalpha(\vec k,t_f) \, 
 e^{- i \omega_k \, (t-t_f)} \, .
 \label{wave}
\end{equation}\noindent
In the following, we drop the implicit dependance on the parameter~$t_f$,
for example,~$\bfalpha(\vec k)$ stands for~$\bfalpha(\vec k,t_f)$
and~$\bfalpha(\vec k,t)$ for~$\bfalpha(\vec k,t>t_f)$.
Moreover, as long as we focus our attention on one particular mode, we shall 
omit the indice~$\vec k$, it will be reintroduced when needed.
The mean number of quanta associated with the classical wave~(\ref{wave}) 
reads 
\beq
 \bar n_j = \langle \alpha | a_j^\dagger a_j  | \alpha \rangle = 
 |\alpha_j (t)|^2 = |\alpha_j|^2
 \label{number}
\end{equation}\noindent
and is time-independant.
The fraction $f$ of neutral pions in the mode $\vec k$ is
\beq
 f = \frac{\bar n_3}{\bar n_1 + \bar n_2 + \bar n_3} \, .
 \label{ratio}
\end{equation}\noindent
\par
Let us now  discuss the polarisation of the ougoing waves~(\ref{wave}).
First, it is easy to see that the motion of the vector 
$\bfvarphi^{out} (t)=\bfvarphi (t>t_f)$ in isospin space is planar. Indeed, 
the vector ${\bf I} = \bfvarphi^{out} \times \dot\bfvarphi^{out}$ is
time-independent\footnote{${\bf I}_{\vec k}$ is the $\vec k$ component of 
  the conserved isorotation generators 
  $\int d^3x \, \bfphi(\vec x,t) \times
  \dot\bfphi(\vec x,t) = \int d^3k \, \bfvarphi(\vec k,t) \times 
  \dot\bfvarphi(\vec k,t)$. For $t>t_f$ the modes are decoupled so each 
  ${\bf I}_{\vec k}$ is conserved.}.
The trajectory described by~$\bfvarphi^{out}$ is an ellipse in the plane 
perpendicular to~${\bf I}$. Let us call~${\bf u}$ and~$L$~(resp.~${\bf v}$
and~$l$) the direction and the half-length of the big axe (resp. of the 
small axe) of this ellipse~(${{\bf u}}^2 = {{\bf v}}^2 = 1$,
${\bf u}.{\bf v} = 0$, ${\bf I} = \omega l L  \, {\bf u} \times {\bf v}$). 
One has then 
\beq
 \bfalpha (t) = \sqrt{\frac{\omega}{2}} 
 \left( L \, {\bf u} + i l \, {\bf v} \right) \, 
 e^{-i ( \omega \, (t-t_f) + \eta)} \, ,
 \label{ellipse}
\end{equation}\noindent
where $\eta$ is some phase factor to be determined from (\ref{wave}).
It follows that
\eq
 \bar n_j = \frac{\omega}{2} \, 
 \left( L^2 \, {u_j}^2 + l^2 \, {v_j}^2 \right) \, ,
\eq
and
\beq
 f = \frac{L^2 \, {u_3}^2 + l^2 \, {v_3}^2}{L^2 + l^2} \, .
 \label{ratio2}
\end{equation}\noindent
\par 
We concentrate on the case of interest, namely the linear polarisation, 
and make explicit the~$\vec k$ dependance.
A linearly polarised wave is characterised by the fact that 
${\bf I}_{\vec k} = {\bf 0}$, that is: $l_{\vec k}=0$. 
Then ($\theta_{\vec k}$ is the angle between 
${\bf u}_{\vec k}$ and the $\pi_3$-axis)
\bearn
 \bfalpha_{linear} (\vec k,t) & = & \alpha (\vec k) \, 
 e^{- i \omega_k \, (t-t_f)} \, {\bf u}_{\vec k} \, , \nonumber \\
 f_{linear}(\vec k) & = & \cos^2 \theta_{\vec k} \, .
 \label{linear}
\eearn
Because both the dynamics (Eq.~(\ref{eom})) and the initial ensemble are
invariants under isospin rotations, there is no privileged direction
in isospin space, so if such a linearly polarised wave is generically
produced, the event-by-event distribution of $f(\vec k)$ will be given by 
the famous $1/\sqrt f$ law. This will remain approximately true for the more 
realistic case of ``flat" elliptic waves ($l_{\vec k} \ll L_{\vec k}$), 
but not otherwise\footnote{It is worth noting that,
  although it is always possible to write $f(\vec k)$ as the squared cosine 
  of some angle $\chi_{\vec k}$, this angle has not, in general, the meaning 
  of a uniform orientation in isospin space so $f(\vec k)$ does not, 
  in general, follow the $1/\sqrt f$ law (see Appendix). In Ref.~\cite{ran}, 
  the author has a $1/\sqrt f$ distribution for $f(\vec k =\vec 0)$, this
  comes from the fact that the time derivative of the field was not included 
  in the definition of the particle number.}.
The event-by-event distribution of the neutral fraction in one mode 
$f(\vec k)$ is then a useful observable and gives a non trivial information 
about the polarisation of the generic outgoing waves $\bfvarphi_{\vec k}^{out}$.
\par
As emphasized in the introduction, in an ideal DCC all 
$\bfvarphi_{\vec k}^{out}$'s oscillate in the same direction 
$\bf u$\footnote{The DCC is a zero isospin state: 
  $\int d^3x \, {\bf I}_{\vec k} = {\bf 0}$~\cite{bj}. This is very explicitly 
  stated in \cite{bjorken}, but is also assumed, more or less explicitly, in 
  all the original papers, where the DCC idea has been put forward.} 
\eq
 \bfalpha_{DCC} (\vec k,t) = \alpha (\vec k) \, 
 e^{- i \omega_k \, (t-t_f)} \, {\bf u} \, .
\eq
So, defining the total neutral pion fraction as
\beq
 f^{tot} =  \frac{N_3}{N_1 + N_2 + N_3} \, ,
 \label{dcc}
\end{equation}\noindent
where $N_j = \int d^3k \, \bar n_j (\vec k)$, one has
\eq
 f_{DCC}^{tot} = \cos^2 \theta \, ,
\eq
where $\theta$ is the angle between $\bf u$ and the $\pi_3$-axis.
\par
For an ideal DCC, the {\em total} neutral pion fraction is distributed
according to the $1/\sqrt f$ law.
In a realistic situation, only those modes undergoing amplification
should be taken into account. Moreover the individual waves are neither
strictly linearly polarised nor strictly aligned with each others, so one
would expect some deviation from the $1/\sqrt f$ law in the distribution
of the total neutral pion fraction $f^{tot}$ (see e.g.~\cite{ran,raj}). 
We will come back to this later. In the present work, we compute the 
following normalized correlation function 
\beq
 C(\vec k,\vec k') = 
 \frac{{\langle f(\vec k) \, f(\vec k') \rangle}_c}
 {\sqrt{{\langle f^2(\vec k)\rangle}_c \, {\langle f^2(\vec k')\rangle}_c}}
  \, ,
 \label{correl}
\end{equation}\noindent
where ${\langle A \, B \rangle}_c = \langle A \, B \rangle - 
\langle A \rangle \, \langle B \rangle$ and~$\langle ... \rangle$ denotes 
the average over the ensemble at time~$t_f$.
In the case of linearly polarised individual waves 
$\bfvarphi_{\vec k}^{out}$, the $f(\vec k)$'s measure the directions 
of polarisation (Eq~(\ref{linear})) and $C(\vec k,\vec k')$ is a measure 
of the degree of alignment of these directions.

\section{Results}

Our box is a $64^3$ cubic lattice and the equations of motion~(\ref{eom}) 
are implemented via a straggered leap-frog scheme~\cite{numrec} with 
time-step~$\Delta t=0.04a$. Once an initial condition has been chosen, 
one can follow the time evolution of the $\bfvarphi(\vec k,t)$s. We 
reproduce completely the results of Ref.~\cite{rw} with the same 
bin-average procedure in momentum space: in the isospin directions~($j=1,2,3$) 
the soft modes experience dramatic amplification, the softer the mode the 
larger the amplification, and exhibit coherent oscillations with
period~$2\pi/\omega_k \sim 2\pi/m_\pi$.
Such behavior does not show up in the~$\sigma$~direction.
This average behavior\footnote{The amplification of a particular mode
   exhibits large fluctuations around its average value over the ensemble. 
   In particular it is not a smooth decreasing function of~$\|\vec k\|$ as 
   revealed by an event-by-event analysis. In this respect, the binning 
   procedure of Ref.~\cite{rw} produces the same result as the ensemble 
   average. The same is observed for the coherent behavior of the soft modes.} 
can be qualitatively understood in the mean-field approximation
\cite{rw,kaiser}: for short times~$t \lesssim 10a$, the curvature of the 
effective potential (effective mass squared) is negative and the softest 
modes, for which the associated effective frequency is imaginary, 
experience amplification: this is the spinodal instability; for 
times~$t \gtrsim 10a$, although this phenomenon is no more efficient 
(the effective mass squared is positive), the field is still significantly 
amplified. This is due to the mechanism of parametric resonance triggered 
by the regular oscillations of the mean-field around its assymptotic 
(vacuum) value~\cite{kaiser,dum}. Obviously, the amplification does not 
grow forever and at late times~($t \gtrsim 100a$) the energy is equally
dissipated among the modes. In the following we present
results for two values of the freeze-out 
time:~$t_f=10a$~($21 \times 10^3$~MC~events), corresponding to the end 
of the spinodal period, and~$t_f=56a$~($10.9 \times 10^3$~MC~events), which 
is the time when the averaged amplification of the soft 
modes~(eq.~(\ref{amplif})) is maximal. For simplicity we only sketch results 
for~$\vec k \equiv (k=n\Delta k,0,0)$,
where~$\Delta k~=~\pi/Na \approx 10$~MeV~($N=64$ is the size of the box). 
Moreover, the linear~$\sigma$-model being an effective
theory for energy scales~$\lesssim 100$ MeV, we only consider the 
window~$0 < n < 15$.
\par
\begin{figure}[p]
%\begin{figure}[h]
\epsfxsize=5.5in \centerline{ \epsfbox{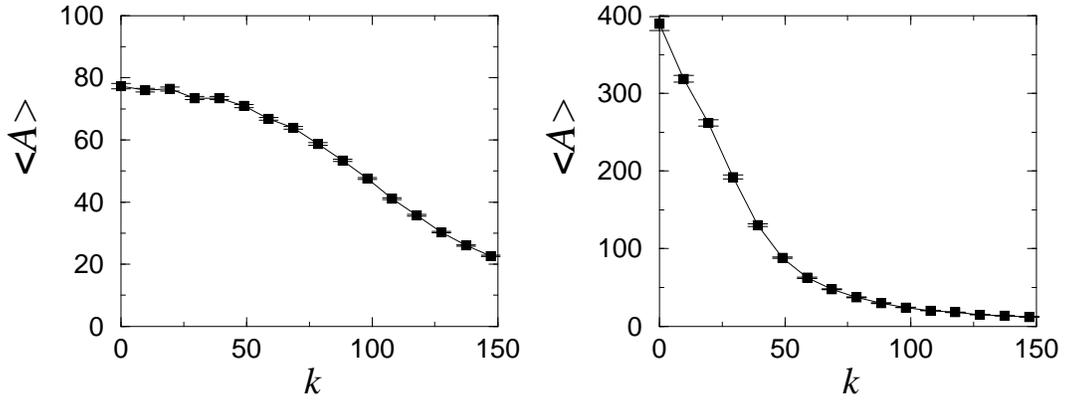}}
\caption[fig1]{\small The average amplification 
$\langle \mathcal A(k,t_f) \rangle$ as a
function of~$k$~(MeV) for~$t_f=10a$~(left) and~$t_f=56a$~(right).
The error bars represent statistical errors and the lines are just 
guides for the eyes.} 
\label{figamplif}
\end{figure}

The amplification factor in the mode~$k$ at time~$t_f$ is defined as
\beq
 \mathcal A(k,t_f) = \frac{P(k,t_f)}{P(k,0)} \, ,
 \label{amplif}
\end{equation}\noindent
where
\eq
 P(k,t_f) = \omega_k \,\sum_{j=1}^3 \, \bar n_j(k,t_f)
\eq
is the power spectrum in the mode $k$ at freeze-out. The mean 
numbers~$\bar n_j(k,t_f)$ are extracted from the field configuration 
at time~$t$ with the help of~Eqs.~(\ref{alpha}) and~(\ref{number}).
The $k$-dependance of the ensemble average 
amplification~$\langle \mathcal A(k,t_f) \rangle$ is shown in 
Fig.~\ref{figamplif}. 
It exhibits the expected shape: low momenta are amplified with respect to 
larger ones. Note the semi-quantitative agreement with mean-field argument 
for~$t_f=10a$: from the mechanism of spinodal instability one expects a 
monotonic enhancement of the modes with momenta~$k \lesssim f_\pi$. 
At~$t_f=56a$ the softest modes have grown further and the window of amplified 
modes has shrunk considerably.
\par
\begin{figure}[p]
%\begin{figure}[h]
\epsfxsize=5.5in \centerline{ \epsfbox{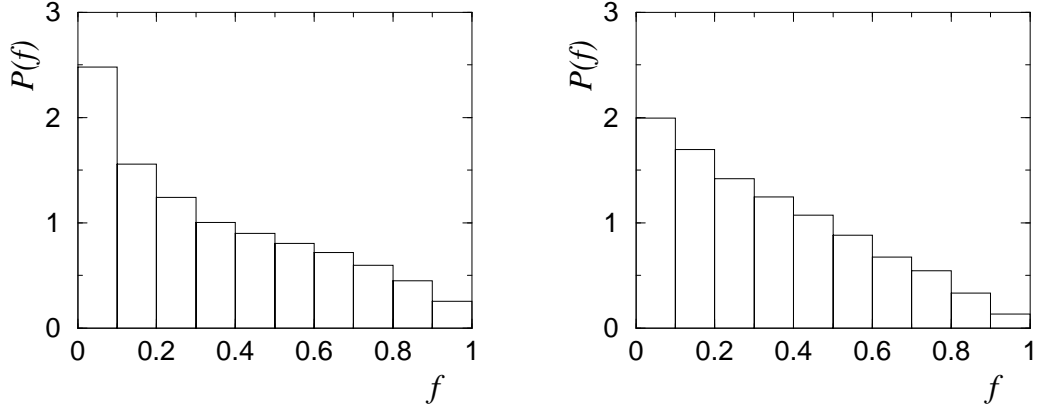}}
\caption[fig2]{\small Neutral fraction distributions at time $t_f=10a$ 
for~$n=0$~(left) and~$n~=~15$~(right)~($k=n\Delta k$ with $\Delta k \simeq 10$~MeV).
All the histograms for~$n \le 14$ are compatible with the corresponding initial
distributions while~$n=15$ already exhibit a linear shape to be compared with
the~$2(1-f)$ neutral fraction distribution in a thermal 
ensemble~(Eq.~(\ref{thermaldist})).}
\label{figpolar1}
\end{figure}
\begin{figure}[p]
%\begin{figure}[h]
\epsfxsize=5.5in \centerline{ \epsfbox{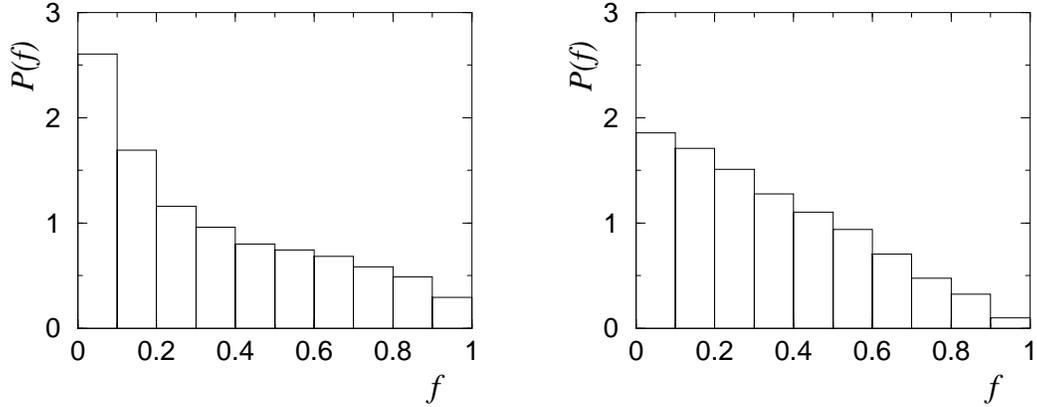}}
\caption[fig3]{\small Neutral fraction distributions at time $t_f=56a$ 
for~$n=0$~(left) and~$n~=~4$~(right)~($k=n\Delta k$ with $\Delta k \simeq 10$~MeV).
The histograms for~$n \le 3$ follow the corresponding initial distributions. 
All the modes~$n \ge 4$ exhibit the same linear shape.} 
\label{figpolar2}
\end{figure}

Event-by-event distributions of the neutral fraction in different modes are
shown for~$t_f=10a$ and~$t_f=56a$ in~Figs.~\ref{figpolar1} and~\ref{figpolar2} 
respectively. All show large fluctuations around the 
mean-value~$\langle f \rangle=1/3$. 
The most amplified modes, those which we are interested in, exhibit a 
very different shape from the less amplified ones. A closer analysis 
reveals that the~$f$-distributions of amplified modes are almost the 
same as in the initial ensemble~(see~(\ref{indist})), whereas those of 
less amplified modes look very different and all follow the same linear 
behavior. It is interesting to note that this linear law is what is 
found in a thermal ensemble of pions~(see~(\ref{thermaldist})), supporting 
the idea that these modes are already thermalised.
\par
\begin{figure}[p]
%\begin{figure}[h]
\epsfxsize=3.5in \centerline{ \epsfbox{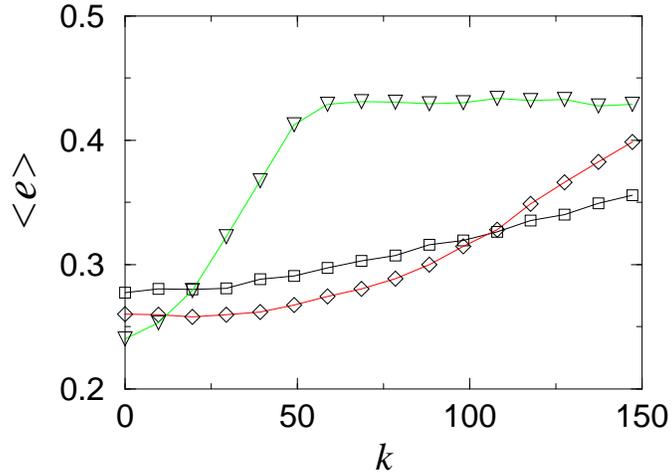}}
\caption[fig4]{\small The average excentricity~$\langle e(k,t_f) \rangle$ as a
function of~$k$~(MeV) in the initial ensemble~(square) and for~$t_f=10a$~(diamond)
and~$t_f=56a$~(triangle). The lines are just guides for the eyes.} 
\label{figellipse}
\end{figure}

Fig.~\ref{figellipse} gives a more precise picture of the above statements.
It represents the average eccentricity\footnote{The eccentricity of an 
   ellipse is defined as the ratio of the lengths of its small to its
   big axes,~i.e., in the notations of~(\ref{ellipse}),~$e_k = l_k/L_k$.} 
of the ellipse~(\ref{ellipse}) as a function of the momentum.
One sees that for~$t_f=10a$ the average polarisation is close to its initial
value for almost the whole studied range of momenta, whereas for~$t_f=56a$
only the few first modes remain close to their initial value, the average
eccentricity being constant for~$k \gtrsim 50$~MeV. Note the correlation 
between the shapes of the average amplification and of the average 
eccentricity as functions of~$k$. Note also that, although the eccentricity 
of the softest modes decreases as these modes are amplified, this is a very 
tiny effect\footnote{An
   event-by-event analysis shows that the proportion of events in which the 
   zero-mode eccentricity is lower than~$0.1$ is:~$13$\% in the initial
   state,~$16$\% at~$t_f=10a$ and~$18$\% at~$t_f=56a$.}
and in fact the shapes of their neutral fraction distribution at freeze-out 
time are mostly determined by the corresponding initial distributions. 
\par
So we see that the modes are far from being strictly linearly polarised 
waves, but this causes only a small a deviation from the ideal~$1/\sqrt f$ 
law. There are still important event-by-event fluctuations of the neutral 
fraction of pions with momentum~$k$, so that from the point of view of an
experimentalist the deviation is not very important.  
\par
\begin{figure}[p]
%\begin{figure}[h]
\epsfxsize=5.5in \centerline{ \epsfbox{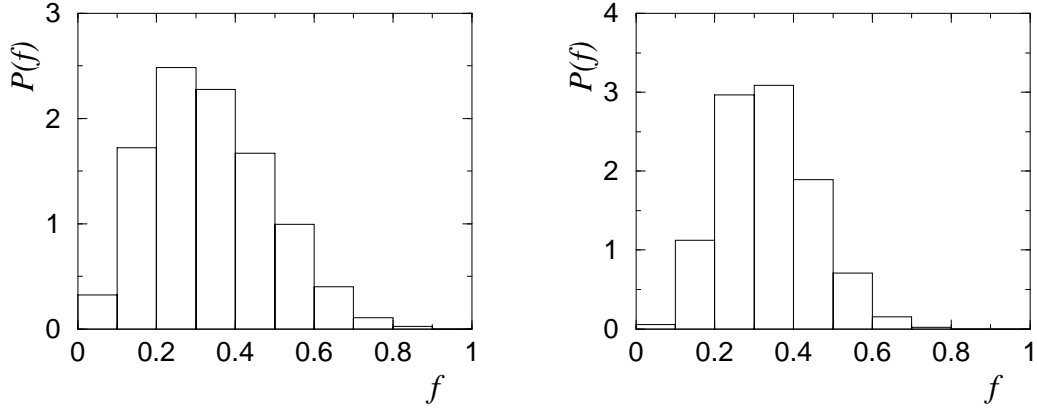}}
\caption[fig6]{\small Neutral fraction distributions where we took into account
the contributions of modes with~$n=0,...,4$~(left) and~$n=0,...,10$~(right) 
at~$t_f=10a$. At this time all these modes have essentially the same individual
distribution~(cf.~Fig.~\ref{figpolar1}).} 
\label{figpolar3}
\end{figure}
\begin{figure}[p]
%\begin{figure}[h]
\epsfxsize=5.5in \centerline{ \epsfbox{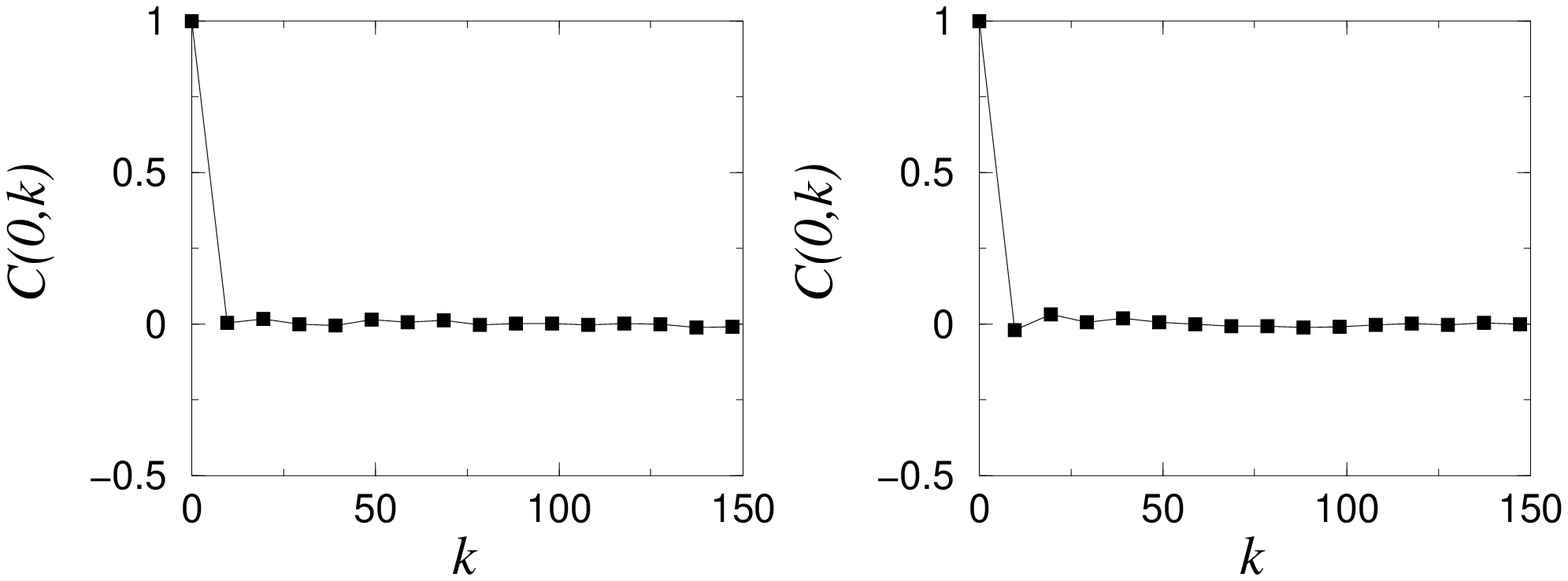}}
\caption[fig5]{\small The correlation function~$C(0,k)$~(Eq.~(\ref{correl}))
vs.~$k$~(MeV) for~$t_f=5a$ (left) and~$t_f=56a$ (right). We computed
also~$C(k_0,k)$ as a function of~$k$ for different values of~$k_0$, 
all look the same as above.} 
\label{figcorrel}
\end{figure}

Finally,~Fig.~\ref{figcorrel} shows the correlation function~(\ref{correl}): 
at both freeze out times the different modes remain completely uncorrelated.
In terms of the average coherent behavior described at the beginning of this 
section, this means that although the vectors~$\bfvarphi_{\vec k}$ 
corresponding to amplified modes do oscillate coherently, their mean 
directions of oscillation in isospin space (say, the directions of 
their big axes) are completely random: different modes act as independant 
DCCs. The obvious phenomenological consequence is that the signal is 
washed out when summing up the contributions of different momenta as in~Eq~(\ref{dcc}). This is illustrated in~Fig.~\ref{figpolar3} where we 
show the distributions coming from the contribution of five and ten 
equivalent modes (that is modes having almost the same individual 
neutral fraction distributions).

\section{Discussion}

\subsection{Summary}

Starting from a chirally symmetric thermal ensemble for the 
field~$\bfphi~\equiv~(\bfpi,\sigma)$, we evolved the system with the
zero temperature equation of motion of the linear~$\sigma$-model.
This quenching of initial (thermal) fluctuations is a very efficient
mechanism for producing large and coherent long wavelength oscillations 
of the pion field at intermediate times~\cite{rw}. 
\par
We performed a detailed analysis of the emerging field configuration 
which shows that nothing significant happens concerning its isospin 
structure. The neutral fraction distribution of the amplified, long 
wavelength modes is essentially the same as it was in the initial ensemble. 
Although not exactly~$1/\sqrt f$, the distribution is very broad, 
which is a relevant point for phenomenology. However, we found that
distinct modes, and in particular the amplified ones, have 
completely independant polarisations in isospin space.
This has the important phenomenological consequence: the large 
event-by-event fluctuations of the neutral fraction are washed out
when the contributions of several modes are added, even when one
limits one's attention to soft modes only.
\par
The key point is to realise that the assumption of a completely
randomized initial state made in~Ref.~\cite{rw} implies that, in the initial
configuration, the field modes are also independant gaussian random numbers.
So our result can be rephrased as follows: the non-linearity of the dynamics
does {\em not} build correlations between modes. The initially postulated
chaos is recovered in the final state. This contradicts the widely made
assumption that the state produced in the simplest form of the quench 
scenario~(initially thermalized system) is identical to the originally proposed 
DCC. The quenching explains the emergence of a strong, coherent pion field, not
its hypothetical polarization.  

\subsection{Comparison with other works}

Analogous results to that shown in~Fig.~\ref{figpolar3} were obtained before 
by several authors~\cite{ggp,raj,ran2}. We briefly review them below.
\par
Gavin, Gocksch and Pisarski argue in~\cite{ggp} that the system is composed of 
``many small, randomly-oriented domains" resulting in a binomial distribution 
of the neutral fraction of the {\em total} number of pions. However, as was 
pointed out later by Rajagopal~\cite{reviews}, the system cannot be 
characterised by a single length scale and the picture of different domains 
of the size of the correlation length is not adequate. In the present paper we 
have shown indeed that the typical field configuration emerging from the 
quench scenario is a superposition of misaligned waves in momentum space:
different {\em modes} and not different domains act as 
independant DCCs\footnote{This does not exclude the possibility
   a priori that many such domains form in heavy ion collisions. However,
   we argued, in a recent work with A. Krzywicki, that the formation of
   many bubbles undergoing strong amplification is not likely~\cite{krzjs}.}. 
\par
In Ref.~\cite{raj}, Rajagopal had in mind a disoriented condensate formed by 
the amplified, long wavelength modes``superposed with short wavelength 
noise". After bining phase space, he computes the bin-by-bin distribution 
of the neutral fraction of pions with~$\| \vec k \| \lesssim 300$~MeV and 
observes ``an admixture of a~$1/\sqrt f$ distribution". This momentum space 
picture is more satisfactory. Indeed the observation that modes of different
types (amplified vs. thermalized) contribute incoherently and delay the 
signal is in line with our present result. However, we saw that soft modes
do not act together to form a coherent misaligned condensate, their 
polarizations are uncorrelated.
\par
Finally, Randrup, studying DCC observables in~\cite{ran2},
computes the distribution of the neutral 
fraction of pions with momenta below different cutoffs. From this he argues
that ``each~$\vec k$ contributes pions having an independant orientation in 
isospace". This is exactly what we have found here. However, Randrup's 
analysis is performed in a model with expansion. This means that not all 
the events entering the analysis have experienced amplification. In this 
sense his work is incomplete and therefore not really conclusive on the 
matter we are interested in here.
\par
Hence, some of the features of the quench scenario have been noticed
in earlier studies. However, nobody has carried out a detailed statistical
analysis of the correlation between mode polarizations. In this respect, 
we believe, our work helps to clarify the situation.

\subsection{Speculations}

As already mentioned, the simplest form of the quench scenario
is not sufficient to produce a DCC configuration. Inclusion of quantum 
effects at the mean-field level or of expansion will probably not 
change this. Instead, it would be instructive to refine the description 
by including correlations in the initial state.  
\par
A possible way
of including initial correlations could be to relax 
the thermal approximation for soft modes.
Indeed, if the quench mechanism appears to be quite natural in the context 
of high energy heavy ion collisions, it is not clear whether the assumption 
that the initial system is fully thermalized is justified. In particular, 
in such small systems, long wavelength modes may not have enough time to 
thermalize, as it is the case in~\cite{rw}~(see also~Fig.\ref{figamplif}).
Correlations in the initial state could possibly be amplified by the 
subsequent out of equilibrium evolution. Indeed, from mean-field
arguments, the spinodal amplification is expected to be quite robust 
against a large class of initial conditions\footnote{The main restriction
   being that initial fluctuations should not be too large in order to enter
   the instability region~(negative effective mass squared).}.
One has then to have a reliable model for the initial state, that is
a model for the early stages of the collision. 
\par
Finally, if some correlations are present in a realistic description
of our initial state, it could be that they do not survive the
non-linear dynamics. One would be left again with an 
incoherent superposition of disoriented waves in the final state.
In such a situation, large fluctuations of the neutral fraction
of pions in a bin of phase space would only be detectable if one
can separate individual modes as well as possible. The volume of the 
bubble should be large enough to have a good statistics in each bin, 
and small enough to have as few as possible modes in each bin.

\subsection{Conclusion}

We believe that a complete microscopic understanding of DCC formation in 
heavy ion collisions is not achieved yet. It might also be, however, that 
the original picture of DCC is a too far going idealization. Future 
investigations will tell whether it is the original picture or the quench 
scenario which has to be modified. Theoretical developpements as well as 
experimental data are called for.

\section*{Aknowledgment}
I am very grateful to A.~Krzywicki, J.P.~Blaizot, and M.~Joyce for useful
conversations and comments and for careful reading of the manuscript.

\appendix
\section{Appendix: neutral fraction distribution in a gaussian model}
\setcounter{equation}{0}
\numberwithin{equation}{section}

 It is instructive to compute the distribution of the neutral fraction~$f$
for the following class of statistical ensembles: let us consider 
the~$\bfvarphi_{\vec k}$'s and the~$\dot \bfvarphi_{\vec k}$'s as independant
gaussian numbers with zero mean value and dispersions~$a$ and~$b$ 
respectively~($a$ and~$b$ are the same for each mode and for each isospin 
direction). We write the probability~($c$ stands for~$\{j,\vec k\}$)
\beq
 Proba \left( \{ \bfvarphi_{\vec k} \},\{\dot \bfvarphi_{\vec k}\} \right) = 
 \prod_c \, P_{\varphi} (\varphi_c) \, P_{\dot\varphi} (\dot\varphi_c) \,
 d\varphi_c \, d\dot\varphi_c \, ,
 \label{ensemble}
\end{equation}\noindent
\bearn
 P_{\varphi} (x) & = & \frac{1}{\sqrt{2 \pi a^2}} \, 
 \exp \left(  - \frac{x^2}{2 a^2} \right) \, , \nonumber\\
 P_{\dot\varphi} (x) & = & \frac{1}{\sqrt{2 \pi b^2}} \, 
 \exp \left(  - \frac{x^2}{2 b^2} \right) \, .
 \label{gaussian}
\eearn
The modes being independant, we focus on one particular $\vec k$ and drop
the indice.
From Eq.~(\ref{alpha}) we have for each isospin component
\bearn
 \varphi & = & \sqrt{\frac{2}{\omega}} \, \mbox{Re} \alpha =
 A \, \cos \gamma  \, ,\nonumber \\
 \dot\varphi & = & \sqrt{2\omega} \, \mbox{Im} \alpha =
 \omega A \, \sin \gamma \, ,
 \label{reim}
\eearn
where $\gamma$ is defined through $\alpha = \sqrt{\bar n} \, e^{i \gamma}$, 
and $A = \sqrt{2 \bar n / \omega}$.
The probability distribution for the amplitude $A$ and the angle $\gamma$ 
is given by
\beq
 P_{A,\gamma} (A,\gamma) = 
 \omega A P_{\varphi} (A \cos \gamma) 
 P_{\dot\varphi} (\omega A \sin \gamma) \, .
 \label{polardist}
\end{equation}\noindent
The probability distribution of the neutral fraction 
$f = {A_3}^2/({A_1}^2 + {A_2}^2 + {A_3}^2)$ is 
\eq
 P_f (f) = \int_0^{+ \infty} dx dy dz \, 
 P_A(x) \, P_A(y) \, P_A(z) \, 
 \delta \left( f - \frac{{z}^2}{x^2 + y^2 + z^2} \right) \, ,
\eq
where 
\eq
 P_A (A) = \int_{0}^{2\pi} d\gamma \, P_{A,\gamma} (A,\gamma) \, .
\eq
After some calculations one gets
\beq
 P_f (f) = \frac{1}{2} \left( F_\Omega (f) + F_{-\Omega} (f) \right) \, ,
 \label{indist}
\end{equation}\noindent
where
\eq
 F_\Omega (f) = \left( \Omega - (1-f) \right) 
 \left( \frac{\Omega + 1}{\Omega - (1-2f)} \right)^{3/2} \, ,
\eq
and
\eq
 \Omega = \frac{b^2 + \omega a^2}{b^2 - \omega a^2} \, .
\eq
\par
It is instructive to look at some specific examples of~(\ref{gaussian}).
First, we can fix one of the two vectors $\bfvarphi$ and $\dot\bfvarphi$
to be zero. In this case $\bfalpha$ is proportional to a randomly oriented 
vector in isospin space and one recovers the~$1/\sqrt f$ law. 
Indeed, let us choose~$b=0$~($\Omega=-1$), that is~$\dot\bfvarphi = {\bf 0}$.
Then~$\bfalpha \sim \bfvarphi$ and
\beq
 P_f (f) = \frac{1}{2 \sqrt f} \, .
\end{equation}\noindent
\par
The second interesting example is the case where $b = \omega a$ 
($\Omega = + \infty$).\\
In terms of a probability distribution for the complex numbers 
$\alpha_j = x_j + iy_j$\footnote{Such a probability distribution for
  the $\alpha$'s is called the P-representation of the density 
  operator~\cite{glauber}: a very large class of quantum density operators
  can be written $\hat \rho = \int d^2 \alpha \, P_{\alpha} (\alpha) \,
  | \alpha \rangle \, \langle \alpha |$.}
\eq
 P_{\alpha} (\alpha) \, d^2\alpha = P_{x,y} (x,y) \, dx dy \, ,
\eq
one has, from~(\ref{gaussian}) and~(\ref{reim}),
\beq
 P_{\alpha} (\alpha) = \frac{1}{\pi \sigma^2} \, e^{-|\alpha|^2/\sigma^2} \, ,
 \label{alphadist}
\end{equation}\noindent
where 
\eq
 \sigma^2 = a\,b = \langle \bar n \rangle = 
 \int d^2\alpha \, P_{\alpha} (\alpha) \, |\alpha|^2
\eq
 ($|\alpha|^2 = \bar n$ is the mean number of quanta in the coherent 
state~$| \alpha \rangle$ and~$\langle \bar n \rangle$ in the mean number 
of quanta in the statistical ensemble).\\
For the class of ensembles~(\ref{alphadist}), the neutral fraction
distribution reads
\beq
 P_f (f) = 2(1-f) \, .
 \label{thermaldist}
\end{equation}\noindent
A particular case of~(\ref{alphadist}) is that of a thermal collection of 
quanta with frequencies $\omega$.
\par
\begin{figure}[p]
%\begin{figure}[h]
\epsfxsize=5.5in \centerline{ \epsfbox{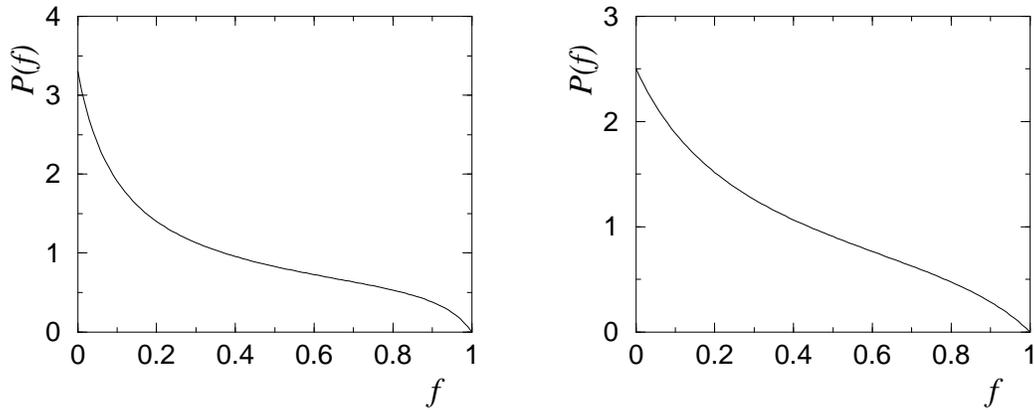}}
\caption[fig4]{\small Neutral fraction distribution in the initial ensemble 
for $n=0$ (left) and $n=15$ (right).} 
\label{figpolar0}
\end{figure}

Finally, the initial ensemble used in the quench scenario belongs to the
class~(\ref{ensemble}). The~$\phi_j (\vec x)$'s and the~$\dot \phi_j (\vec x)$'s 
are independant gaussian numbers of variances $\langle \phi_j^2 \rangle$ and 
$\langle \dot\phi_j^2 \rangle$ respectively. This implies that 
the~$\varphi_j (\vec k)$'s and the~$\dot \varphi_j (\vec k)$'s are independant
gaussian numbers of variances~${\cal N} \, \langle \phi_j^2 \rangle$ 
and~${\cal N} \, \langle \dot\phi_j^2 \rangle$ respectively ($\cal N$ is a 
normalisation factor)
\eq
 a^2 = {\cal N} \, \langle \phi_j^2 \rangle = 
 {\cal N} \, \frac{v^2}{16} \, , \, \,
 b^2 = {\cal N} \, \langle \dot\phi_j^2 \rangle = 
 {\cal N} \, \frac{v^2}{4} \, ,
\eq
and
\eq
 \Omega = \frac{4 + \omega^2}{4 - \omega^2} \, .
\eq
Fig.~\ref{figpolar0} shows the neutral fraction distribution in the
initial ensemble for the two extremal values of the studied 
window~$0 \leq k \lesssim 150$ MeV\footnote{Note that although our initial 
state is a thermal collection of quanta, these are not zero temperature
pions (this means that their dispersion relations are not those of
free pions). That is why the initial neutral fraction distribution
is not the same as~(\ref{thermaldist}).}.

%%%%%%%%%%%%%%%%%%%%%%%%%%%%%%%%%%%%%%%%%%%%%%%%%%%%%%%%%%%%%%%%%%%%%%%%%%%
%%%%%%%%%%%%%%%%%%%%%%%%%%%%%%%%%%%%%%%%%%%%%%%%%%%%%%%%%%%%%%%%%%%%%%%%%%%
%%%%%%   l'option [h] permet de placer la figure la ou on la veut.  %%%%%%%
%%%%%%   l'option [p] (temporaire) met toute les figures a part     %%%%%%%
%%%%%%%%%%%%%%%%%%%%%%%%%%%%%%%%%%%%%%%%%%%%%%%%%%%%%%%%%%%%%%%%%%%%%%%%%%%
%%%%%%%%%%%%%%%%%%%%%%%%%%%%%%%%%%%%%%%%%%%%%%%%%%%%%%%%%%%%%%%%%%%%%%%%%%%

\newpage
\begin{thebibliography}{99}
 \bibitem{bjorken} G. Amelino-Camelia, J.D. Bjorken, S.E. Larsson
Phys. Rev. D56 (1997) 6942.
 \bibitem{reviews} K. Rajagopal, in {\em Quark-Gluon Plasma 2}, 
ed. R. Hwa, World Scientific, 1995.;\\
J.P. Blaizot and A. Krzywicki, 
Acta Phys. Polon. 27 (1996) 1687.
 \bibitem{exp} The Minimax Collaboration, 
Phys. Rev. D61 (2000) 032003;
The WA98 Collaboration,
Nucl. Phys. A638 (1998) 249c.
 \bibitem{bj} J.D. Bjorken, 
Int. J. Mod. Phys. A7 (1992) 4189; 
Acta Phys. Polon. B23 (1992) 637.
 \bibitem{bk} J.P. Blaizot and A. Krzywicki, 
Phys. Rev. D46 (1992) 246;
Phys. Rev. D50 (1994) 442.
 \bibitem{rw} K. Rajagopal and F. Wilczek, 
Nucl. Phys. B399 (1993) 395; 
Nucl. Phys. B404 (1993) 577.
 \bibitem{boy} D. Boyanovsky, H.J. de Vega, R. Holman
Phys. Rev. D51 (1995) 734.
 \bibitem{cooper} F. Cooper, Y. Kluger, E. Mottola, J.P. Paz,
Phys. Rev. D51 (1995) 2377; 
M.A. Lampert, J.F. Dawson, F. Cooper,
Phys. Rev. D54 (1996) 2213.
 \bibitem{ran} J. Randrup,
Phys. Rev. Lett. 77 (1996) 1226.
 \bibitem{ggp}  S. Gavin, A. Gocksch, R.D. Pisarski
Phys. Rev. Lett. 72 (1994) 2143.
 \bibitem{raj} K. Rajagopal, hep-ph/9703258
 \bibitem{ran2} J. Randrup,
Nucl. Phys. A616 (1997) 531. 
 \bibitem{krzjs} A. Krzywicki and J. Serreau, 
Phys. Lett, B448 (1999) 257.
 \bibitem{numrec} Numerical Recipes, 
ed. Cambridge University Press.
 \bibitem{kaiser} D. Kaiser,
Phys. Rev. D59 (1999) 117901.
 \bibitem{dum} A. Dumitru and O. Scavenius,
hep-ph/0003134.
 \bibitem{glauber} R.J. Glauber,
Phys. Rev. 131 (1963) 2766.
\end{thebibliography}
\end{document}